# An IoT-based Smart Parking System

Ridhi Choudhary, Arnav Sanjay Sinha, Krishna Jaiswal, Anurag Chandra

Vellore Institute Of Technology

Chennai, India

**Abstract -** The number of vehicles on the road is growing every day, thus there's a growing need to develop effective and hassle-free parking systems. Finding a parking space may be a big challenge, especially in crowded cities or areas with scheduled sporting or cultural events. The project suggests an automated parking system that makes use of technology like sensor systems and microcontrollers. In order to make it easier for drivers to park in empty spots and cut down on the time and effort needed for manual searches, this system is made to identify empty parking spaces and display the available parking spots on an LCD screen.

## I. INTRODUCTION

Finding a parking spot in busy city streets can be a frustrating experience due to limited parking spaces. To address this issue, we present a smart parking system that uses Arduino, an open-source microcontroller platform. Our system aims to optimize parking by monitoring the availability of parking spaces in real-time and providing valuable information to drivers.

The system consists of sensors installed in each parking slot, which detect the presence of vehicles. These sensors are connected to an Arduino microcontroller, which processes the sensor data and determines whether a parking space is occupied or vacant. The status of each parking space is then displayed on a screen, allowing drivers to quickly identify available spots.

Additionally, the system features a barrier controlled by a motor. The barrier opens and closes based on the availability of parking spaces. When all the spots are taken, the barrier restricts entry to prevent unnecessary traffic congestion within the parking area. To enable real-time monitoring and management, the Arduino system is equipped with wireless communication modules. This allows the system to transmit parking occupancy data to a central server or a mobile application. Drivers can access this information remotely, making it easier for them to find available parking spaces. The benefits of this smart parking system are significant. Drivers experience reduced search time, making their parking experience more convenient. Authorities can efficiently manage parking resources, enforce regulations effectively, and gather valuable data for future planning.

## II. LITERATURE REVIEW

In [1] proposed a parking spot identification system using the turn detection mechanism. A Smart Parking System using Low Energy BlueTooth Beacons With Particle Filtering was proposed by [2]. [3] proposed the deployment of WSN for Smart Car Parking Systems both in indoor and outdoor environments. [4] proposed An Intelligent Real-Time Parking Monitoring and Automatic Billing System using IoT. [5] proposed the Mobile Outdoor Parking Space Detection Application using the detection mechanism. Survey for a Cloud-Based Parking System using IoT Technology by [6]. Investigation of Smart Parking Systems and their technologies by [7].

## III. METHODOLOGY:-

1. Job Description:
About the requirements and operation of the station.
Set the number of parking spaces.

2. System Architecture:
Design the entire architecture of the station. Install necessary components such as Arduino board, LCD screen, IR sensor and other accessories.

3. Hardware installation:
Connect the Arduino board to the components (LCD display, infrared sensor).
Make sure the connection is correct and secure.
Test products one by one to make sure they work as intended.

4. Arduino Programming:
Write Arduino code to control and manage the system.
According to infrared sensor data processing logic to detect the vehicle.
Create a function to update the LCD screen based on sensor input.
The code that the user enters for a reason such as saving or leaving the parking space.

5. User Interface (LCD):
Create a user interface on the LCD screen.
Show status information for all stots (full or not).
Make the interface user-friendly and intuitive.
.
6. Testing:
Test all systems in a controlled environment.
Simulate vehicle conditionally and unconditionally to check sensor accuracy.
Check that the LCD screen is adjusted correctly according to the sensor input.

7. Integration:
Integrate all components and test the system as a whole.
Ensure good communication between Arduino, sensor and LCD screen.

8. Information:
List hardware connections and settings.
Provide detailed instructions on the use of codes.
Contains a user guide explaining how to interact with the slots.

*Data Set Description:-*

Parking Slot Status:

Description: Represents the status of each parking slot (occupied or vacant).
Data Type:String(Fill or Empty)
Usage: This data is used to determine the availability of parking slots. It is updated based on the input from the IR sensors.
IR Sensor Data:

Description: Raw data from the IR sensors indicating the presence or absence of a car in a specific parking slot.
Data Type: int(0 or 1).

Usage: The Arduino processes this data to determine the occupancy status of each parking slot. Calibration parameters may be included to optimise sensor performance.

LCD Display Data:

Description: Data sent to the LCD screen for displaying parking slot status and user instructions.
Data Type: String or custom data structure.
Usage: The Arduino generates this data based on the parking slot status. It includes information such as slot numbers, availability.

*Models:-*

1. Parking Slot Status Model:

Description: The status of each parking slot is a binary value indicating whether the slot is occupied or vacant.

Mathematical Equation:

1 -> if car is detected in slot
0 -> if slot is vacant

2. IR Sensor Model:

Description: The IR sensors generate binary data indicating the presence or absence of a car in a specific parking slot.

Mathematical Equation:

1 -> if car is detected in slot
0 -> if slot is vacant

3. LCD Display Model:

Description: The LCD display shows the status of each parking slot based on the information from the Parking Slot Status model.

Mathematical Equation:
total=s1+s2+s3+s4
slot = slot + total

IV.CONCLUSION AND FUTURE ASPECT
Our Vehicle Parking Management System is a solution to the challenges of managing parking in big cities where parking space is limited and monitoring of vehicles is difficult. This program is designed to be efficient and can be managed by a single person. We've achieved many of our goals but are still working on improving the system by developing methods for obtaining voltage using piezoelectricity. We would appreciate your honest feedback to help us make this hardware even more efficient and add new features. Thank you for your support!

V. RESULT

We monitored the presence of vehicles in real-time using sensors and analysed the project to make parking easier for everyone. Our proposed solution is designed to be user-friendly, with essential features that are easy to use before parking a vehicle. Although there is still room for improvement, we have implemented our project as planned and are excited about the potential it has to become the perfect application for a vehicle parking system.